\documentclass[
 reprint,
 amsmath,
 amssymb,
 superscriptaddress,
 aps,
]{revtex4-1}

\usepackage{graphicx}
\usepackage{bm}
\usepackage{dcolumn}
\usepackage{gensymb}
\usepackage[utf8]{inputenc}
\usepackage{float}

\usepackage{array,mathtools,amssymb,booktabs}
\newcolumntype{C}{>{$}c<{$}}
\AtBeginDocument{
\heavyrulewidth=.08em
\lightrulewidth=.05em
\cmidrulewidth=.03em
\belowrulesep=.65ex
\belowbottomsep=0pt
\aboverulesep=.4ex
\abovetopsep=0pt
\cmidrulesep=\doublerulesep
\cmidrulekern=.5em
\defaultaddspace=.5em
}

\usepackage{hyperref}
\usepackage{siunitx}
\newcommand{\tref}[1]{Table~\ref{#1}}

\begin{document}

%\preprint{APS/123-QED}

\title{Measurement of the $7p\,\, ^2\!P_{3/2}$ state branching fractions in $\mathrm{Ra}^+$}

\author{M. Fan}
\affiliation{Department of Physics, University of California, Santa Barbara, California 93106, USA}
\affiliation{California Institute for Quantum Entanglement, Santa Barbara, California 93106, USA}
\author{C. A. Holliman}
\affiliation{Department of Physics, University of California, Santa Barbara, California 93106, USA}
\affiliation{California Institute for Quantum Entanglement, Santa Barbara, California 93106, USA}
\author{S. G. Porsev}
\affiliation{Department of Physics and Astronomy, University of Delaware, Newark, Delaware 19716, USA}
\affiliation{Petersburg Nuclear Physics Institute of NRC ``Kurchatov Institute'', Gatchina, Leningrad District 188300, Russia}
\author{M. S. Safronova}
\affiliation{Department of Physics and Astronomy, University of Delaware, Newark, Delaware 19716, USA}
\affiliation{Joint Quantum Institute, National Institute of Standards and Technology and the University of Maryland, College Park, Maryland 20742, USA}
\author{A. M. Jayich}
\email{jayich@gmail.com}
\affiliation{Department of Physics, University of California, Santa Barbara, California 93106, USA}
\affiliation{California Institute for Quantum Entanglement, Santa Barbara, California 93106, USA}

\begin{abstract}
We report a measurement of the radium ion's $7p\,\, ^2\!P_{3/2}$ state branching fractions and improved theoretical calculations. With a single laser-cooled $^{226}\mathrm{Ra}^+$ ion we measure the $P_{3/2}$ branching fractions to the $7s\,\,^2\!S_{1/2}$ ground state $\SI{0.87678\pm0.00020}{}$,
the $6d\,\,^2\!D_{5/2}$ state $\SI{0.10759\pm0.00010}{}$, and the $6d$ ${}^{2}D_{3/2}$ state $\SI{0.01563\pm0.00021}{}$.
\end{abstract}

\date{\today}

\maketitle

% ====================
\section{Introduction}
% ====================
Precise values for electric dipole matrix elements (MEs) provide fundamental knowledge for atomic and molecular systems and are needed for many applications, including studies of fundamental symmetries and development of atomic clocks. Precision measurements are vital for development of high-precision theory, in particular for heavy systems. For example, in an atomic parity non-conservation (PNC) experiment precise information about the atom's electronic structure is critical to compare the experimental result with the prediction of the standard model. A single radium ion has been considered for PNC measurements due to both its large nuclear charge ($Z=88$), as PNC effects scale as $Z^3$, and the high degree of control available in the
system \cite{Fortson1993, Bouchiat1997, Geetha1998, DzuFlaGin01, Mandal2010}.

For an electronic state connected to multiple lower-lying states through dipole allowed transitions an extraction of MEs from the lifetime measurements requires measuring the corresponding branching fractions. Here we report a measurement of the radium ion's $7p\,\, ^2\!P_{3/2}$ branching fractions to the ground $7s\,\,^2\!S_{1/2}$ state and the long-lived $6d\,\,^2\!D_{3/2}$ and $6d\,\,^2\!D_{5/2}$ states.

The $E1$ transition amplitudes were calculated earlier for a number of the low-lying states, using different methods~\cite{Pal2009, Sahoo2007, RobDzuFla13a}. In particular, in Ref.~\cite{Pal2009} the calculations were carried out by the all-order method including single double excitations (SD) and perturbative triple excitations. All non-linear terms and non-perturbative triples were omitted in~\cite{Pal2009}.
The SD approach is equivalent to a linearized coupled-cluster single double (LCCSD) method.

In this paper we carried out calculations in the framework of the LCCSD method and also included full valence triples excitations (solving the equations for triple cluster amplitudes iteratively) and  non-linear terms. Based on Cs high-precision studies~\cite{PorBelDer10} we can expect strong cancellation of these contributions, but very few experimental results are of sufficient accuracy to allow a comprehensive assessment of these effects, and of these measurements most are in lighter systems. This paper provides needed benchmarks to gauge the importance of these effects for  heavy atoms.

Moreover, using very precise measurement of the $P_{3/2}$ to $D_{5/2}$ branching fraction and an accurate calculation of the ratio of the $P_{3/2}$ to $D_{3/2}$ and $P_{3/2}$ to $D_{5/2}$ branching fractions, we are able to extract the value of the $P_{3/2}$ to $D_{3/2}$ branching fraction, reducing its uncertainty by a factor of 2 compared to the pure experimental result.

The theory-experimental comparison carried out here also provides important information for predicting properties of superheavy elements with $Z > 100$ where precision theory is needed for prediction of energies and matrix elements prior to difficult one-atom-at-a-time spectroscopy studies~\cite{PorSafSaf18}.  Precision theory predictions allow for quick transition searches, which are particularly important due to limited beam time.

\section{Experimental Setup}

\begin{figure}
    \centering
    \includegraphics[width=0.75\linewidth]{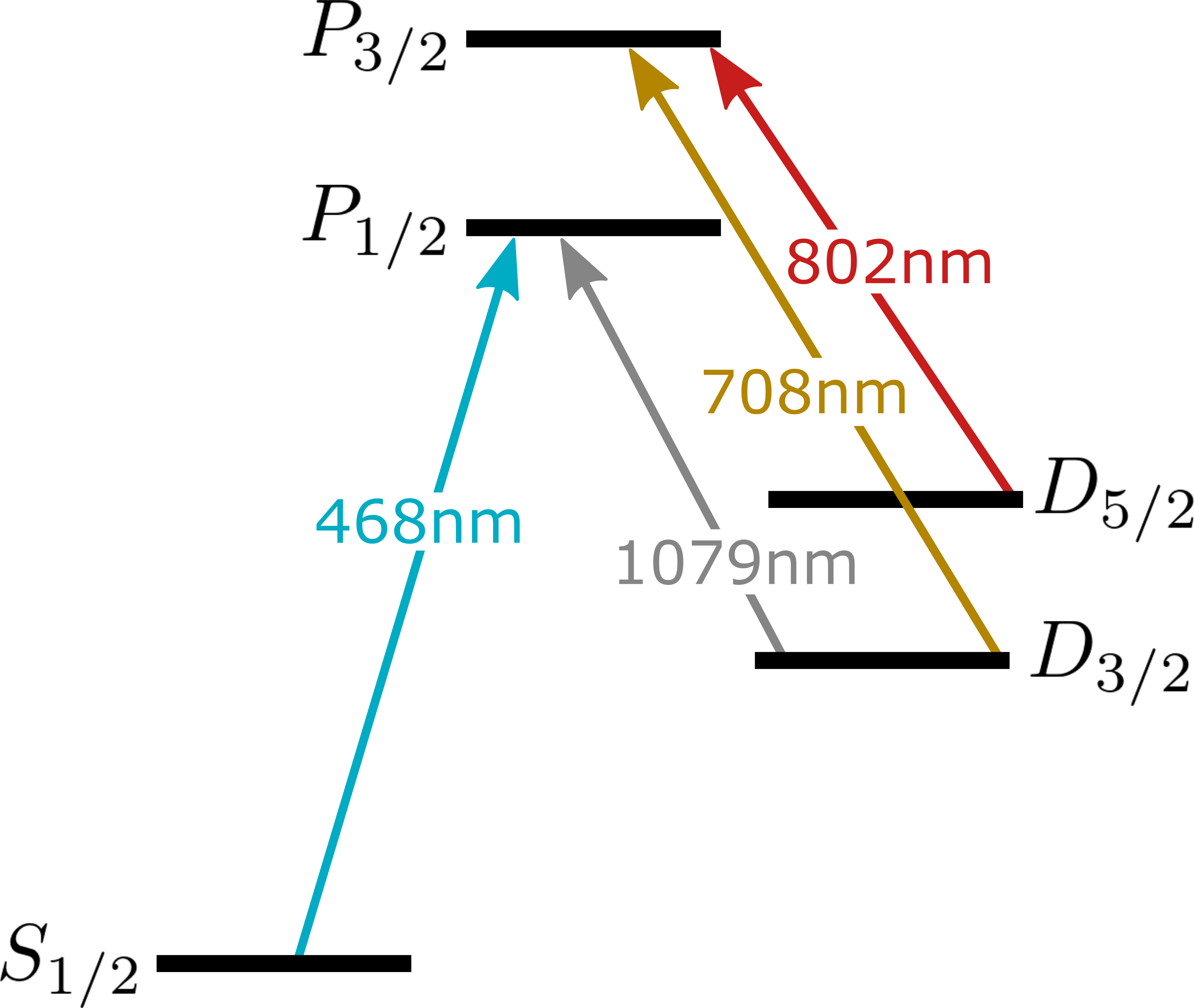}
    \caption{The laser wavelengths and radium ion energy levels used to measure the $P_{3/2}$ branching fractions.}
    \label{fig:energy_structure}
\end{figure}

We measure the branching fractions of the $P_{3/2}$ state to the $S_{1/2}$ ($r$), $D_{5/2}$ ($s$), and $D_{3/2}$ ($t$) states using a single laser-cooled $^{226}\mathrm{Ra}^+$ ion in a linear Paul trap. The relevant energy levels and laser wavelengths are shown in Fig. \ref{fig:energy_structure}. The experimental setup is described in \cite{Fan2019}.  In this paper the rf trapping frequency is \SI{1.8}{\mega\hertz} and a static magnetic field of about 3 G is applied along the trap's axial direction.  Similar precision measurements of branching fractions from the $P_{3/2}$ state have been done in $\mathrm{Ca}^+$ \cite{Gerritsma2008}, $\mathrm{Sr}^+$ \cite{Zhang2016}, and $\mathrm{Ba}^+$ \cite{Dutta2016}.

\begin{figure}[H]
    \centering
    \includegraphics[]{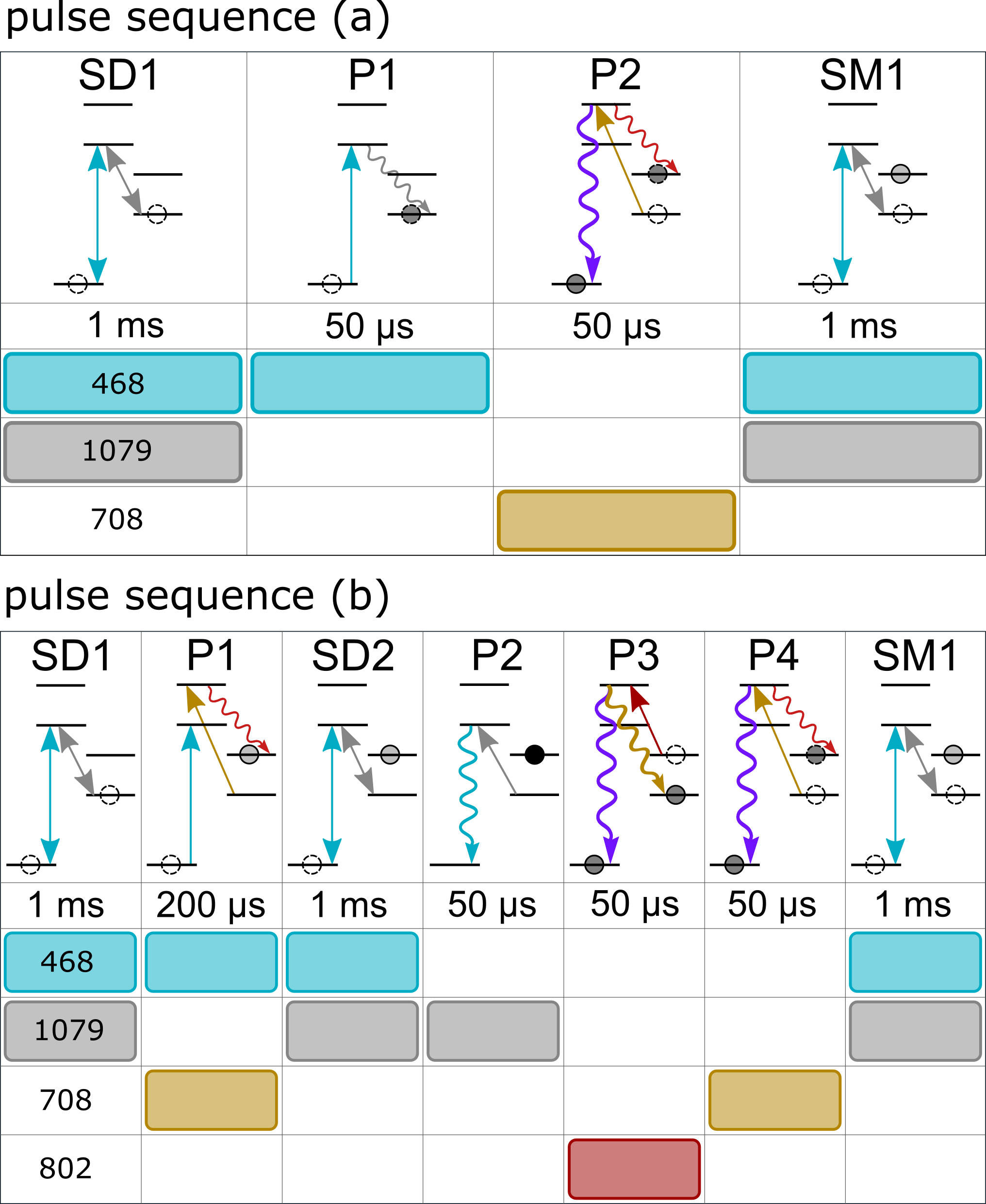}
    \caption{The pulse sequences (a) and (b) for measuring the radium ion's $P_{3/2}$ state branching fractions.  The abbreviated energy level structure is shown in detail in Fig. \ref{fig:energy_structure}. Each sequence is preceded by \SI{50}{\micro\second} of 802 nm cleanout from the $D_{5/2}$ state and \SI{200}{\micro\second} of Doppler cooling. Sequence (a) is repeated \num{11360000} times, and sequence (b) is repeated \num{3050000} times.}
    \label{fig:sequence}
\end{figure}

All laser frequencies and amplitudes for cooling and optical pumping are controlled with double-pass acousto-optic modulators (AOMs). We program pulse sequences to a field-programmable gate array that controls the AOMs \cite{Pruttivarasin2015}. Because the $P_{3/2}$ state decays to three states we use two pulse sequences, labeled as pulse sequence (a) and pulse sequence (b) in Fig. \ref{fig:sequence}, to measure the three branching fractions.

In both sequences we perform state detection where 468 nm light is collected on a photomultiplier tube (PMT) while the $S_{1/2}-P_{1/2}$ and $D_{3/2}-P_{1/2}$ transitions are driven at 468 and  1079 nm, respectively.  If the ion fluoresces, the population was in the $S_{1/2}$ and $D_{3/2}$ states (bright states), and we denote the state detection as a bright event.  If the ion does not fluoresce, the ion was either shelved in the $D_{5/2}$ state (dark state) or has left the imaging region, and we denote the state detection as a dark event.  During 1 ms of state detection, if the ion was in the $S_{1/2}$ and $D_{3/2}$ states we count on average 35 photons with a PMT, whereas if the population was in the $D_{5/2}$ state there is only one count on average.  We set a state detection threshold at 10.5 counts, which detects bright events with greater than \num{99.997}\% efficiency from Poisson statistics. However, due to the $D_{5/2}$ state decays,  \num{0.2}\% of dark events where the population starts in the $D_{5/2}$ state at the beginning of state detection are mislabelled as bright events with the state detection method.

\begin{figure}
    \centering
    \includegraphics[]{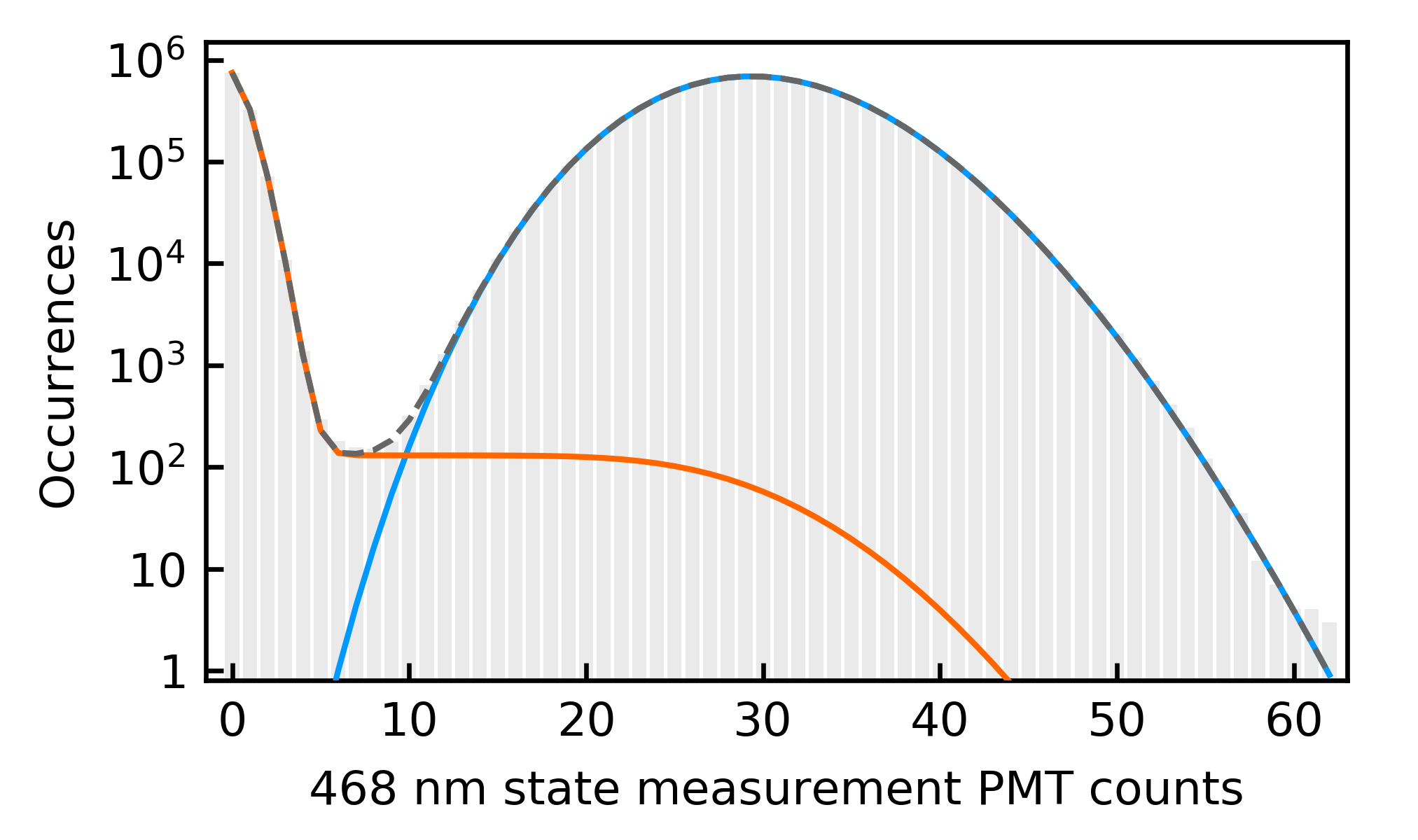}
    \caption{A histogram of 468 nm PMT counts during state measurement 1 (SM1) of sequence (a). The $x$ axis shows PMT counts in 1-ms state measurements, and the $y$ axis shows the occurrences of each PMT count. The maximum likelihood calculation for the dark state probability yields $p_{a}=0.10928(10)$ for sequence (a) measurement. The orange curve shows the PMT counts distribution of dark events, and the blue curve shows the PMT counts distribution for bright events. The gray dashed curve shows the PMT counts distribution for all events, the sum of dark and bright.}
    \label{fig:histogram}
\end{figure}

We also perform state measurements, where again 468 and 1079 nm light is used, but apply a maximum likelihood technique that analyzes the PMT counts from all measurements to calculate the $D_{5/2}$ state population \cite{Gerritsma2008}. We model the bright state counts as a Poisson distribution. The dark state counts are modeled as a weighted sum of Poisson distributions where the average dark state counts increase if the $D_{5/2}$ state decays closer to the start of the 1-ms state measurement pulse [see Fig. \ref{fig:sequence} pulse sequences (a) and (b) SM1].  We use the theoretical lifetime $303(4)$ ms from \cite{Pal2009} to calculate the dark state Poisson distribution weights. The dark events occur with probability $p_\mathrm{d}$ (bright events occur with probability $p_\mathrm{b}=1-p_\mathrm{d}$).  The $p_\mathrm{d}$ value maximizing the probability that the experimentally collected counts are observed (see Fig. \ref{fig:histogram})  is the maximum likelihood value.  The uncertainty of $p_\mathrm{d}$ is $\Delta p_\mathrm{d}=\sqrt{p_\mathrm{d}(1-p_\mathrm{d})/M}$, where $M$ is the number of state measurements.  For more information on the state measurement PMT counts model and the maximum likelihood technique see Appendix A.

Pulse sequences (a) and (b) in Fig. \ref{fig:sequence} begin with 1 ms of state detection.  At the end of both pulse sequences we optically pump at 802 nm for \SI{50}{\micro\second} to remove population from the $D_{5/2}$ state, and then laser cool for \SI{200}{\micro\second}.

The pulse sequence (a), Fig. \ref{fig:sequence}, measures the ratio of the $P_{3/2}$ branching fractions to the $S_{1/2}$ and $D_{5/2}$ states.  After the initial state detection (SD1) the population is optically pumped for \SI{50}{\micro\second} to the $D_{3/2}$ state with 468 nm light (P1).  The population is then pumped at 708 nm for \SI{50}{\micro\second} through the short-lived $P_{3/2}$  state to the $S_{1/2}$ and $D_{5/2}$ states (P2).  Then 1 ms of state measurement (SM1) determines whether the ion is in the $S_{1/2}$ or the $D_{5/2}$ state.  The measured $D_{5/2}$ population fraction of SM1, $p_{a}$, is related to the branching fractions $r$ and $s$ by $p_{a} = s / (r + s)$.

The pulse sequence (b), Fig. \ref{fig:sequence}, does not measure a simple quantity, such as a branching fraction ratio, but when combined with the sequence (a) result we can determine all of the $P_{3/2}$ branching fractions.  After the initial state detection the ion is optically pumped to the $D_{5/2}$ state using \SI{200}{\micro\second} of 468 and 708 nm light (P1). We state detect for 1 ms to verify pumping to the $D_{5/2}$ state (SD2).  Population that might have entered the $D_{3/2}$ state during SD2 is optically pumped to the ground state  with a \SI{50}{\micro\second} 1079 nm pulse (P2).  The $D_{5/2}$ population is then optically pumped (P3) with 802 nm light (\SI{50}{\micro\second}) through the $P_{3/2}$ state to populate the $S_{1/2}$ and $D_{3/2}$ states.  The $D_{3/2}$ population is then pumped at 708 nm  (\SI{50}{\micro\second}, P4) through the $P_{3/2}$ state the decay of which increases the ground state and the $D_{5/2}$ state populations.  The $D_{5/2}$ state population fraction,  $p_{b}$, is measured with a final state measurement (1 ms, SM1).  The relationship between the measured $D_{5/2}$ state population fraction, $p_{b}$, and the branching fractions
is $p_{b} = p_{a}\times t / (r+t)$.

\section{Experimental results}

We condition the data from both pulse sequences based on the state detection results. For sequence (a) if the first state detection after Doppler cooling is dark, we reject the data point as either the ion was not in the imaging region or the ion was shelved in the long-lived $D_{5/2}$ state. From \num{11360000} events \num{753482} were rejected, where the majority of rejected events were excluded due to the electron population being shelved in the $D_{5/2}$ state at the start of the sequence. The rejected events stem from the fact that 802 nm light is not used during sequence (a), and therefore is not kept on resonance so occasionally the 802 nm reset pulse at the end of each sequence may fail.

Similarly, for sequence (b) we discard the data point if the first state detection is dark (\num{962} rejected out of \num{3050000}). We keep 802 nm light on resonance with the $\mathrm{Ra}^+$ $D_{5/2}- P_{3/2}$ transition for sequence (b), which results in fewer rejected SD1 data points compared to sequence (a). We count consecutive rejected data points as a single collision event, where the ion either gains sufficient kinetic energy to leave the imaging region or is shelved to the $D_{5/2}$ dark state due to an inelastic collision. The rejected data correspond to 19 collision events, yielding a collision rate of 0.0018 Hz, agreeing with the measured collision rate of \SI{0.0017\pm0.0004}{\hertz} (See Appendix B). If the second state detection is bright, we also discard the data point, as it indicates that the P1 pumping step failed, or the ion decayed from the $D_{5/2}$ state. Of \num{3049038} events \num{8049} were rejected, which agrees with the decay probability from the $D_{5/2}$ state during state detection given the $D_{5/2}$ lifetime and $D_{5/2}$ depopulation rate due to 802 nm AOM leakthrough (Appendix C).

With the equations for $p_{a}$ and $p_{b}$, we can calculate the branching fractions

\begin{align}
    r & = \frac{(1-p_{a})(p_{a}-p_{b})}{p_{a}(1-p_{b})} , \\
    s & = \frac{p_{a}-p_{b}}{1-p_{b}},  \\
    t & = 1-r-s .
\end{align}

We have \num{10606518} data points for sequence (a), and the maximum likelihood value for the dark event probability is $p_{a}=\SI{0.10928\pm0.00010}{}$. We have \num{3040989} data points for sequence (b), and the maximum likelihood value for the dark event probability is $p_{b}=\SI{0.00192\pm0.00003}{}$.  Using Eqs.\,(1--3), we calculate the statistical branching fractions: $r_{\mathrm{stat}} = \SI{0.87677\pm0.00020}{}$, $s_{\mathrm{stat}} = \SI{0.10757\pm0.00010}{}$, and $t_{\mathrm{stat}} = \SI{0.01565\pm0.00021}{}$.
%---------------------------
\section{Systematic effects}
%---------------------------
The systematic uncertainties and shifts that affect the branching fractions along with the statistical results are summarized in Table \ref{table:branching_fraction_systematics}.  The reported uncertainties represent one standard deviation. The systematics are discussed below with further details in the Appendices B-D.

\begin{table*} \centering
\caption{Shifts and uncertainties for the $P_{3/2}$ branching measurement.
\label{table:branching_fraction_systematics}}
\begin{ruledtabular}
\begin{tabular}{lccc}
%\rowcolor{gray!50}
    Source & $r$ & $s$ & $t$  \\ 
    \midrule
    Statistical & 0.87677(20) & 0.10757(10) & 0.01565(21) \\
    Collisions & $\SI{0\pm4e-5}{}$ & $0.0(1.1)\times10^{-5}$ & $\SI{0\pm4e-5}{}$ \\
    802 nm AOM Leak SM & $-2.7(1.0)\times10^{-5}$ & $2.7(1.1)\times10^{-5}$ & $\SI{0\pm3e-6}{}$ \\
    State Detection Fidelity & $\SI{1.5\pm0.4e-5}{}$ & $1.8(1.1)\times10^{-6}$ & $\SI{-1.6\pm0.4e-5}{}$ \\
    Finite $D_{5/2}$ and $D_{3/2}$ Lifetimes & $\SI{1.48\pm0.02e-5}{}$ & $\SI{-1.117\pm0.019e-5}{}$  & $\SI{-3.63\pm0.11e-6}{}$ \\
    AOM Leak State Preparation & $2.6(1.5)\times10^{-6}$ & $-2.5(1.2)\times10^{-6}$ & $\SI{-1\pm7e-7}{}$ \\
    \bottomrule \\[-7pt]
    \bf{Total} & 0.87678(20) & 0.10759(10) & 0.01563(21) \\ 
\end{tabular}
\end{ruledtabular}
\end{table*}

There is a systematic uncertainty due to collisions.  Inelastic collisions can change the ion's electronic state. Elastic collisions can Doppler shift the ion's transitions or bump the ion out of the imaging region, reducing the number of scattered photons and leading to a false dark detection event when the ion is in a bright state. Both inelastic and elastic collisions shift the state measurement probabilities. Electric-field noise may also transfer kinetic energy to the ion and shift the state measurement probabilities, but the shift is small compared to collisions in our setup. We measure a total collision rate of \SI{0.0017\pm0.0004}{\hertz} for a single radium ion in the trap (Appendix B). We assume a maximum collision rate of \SI{0.0021}{\hertz} to calculate shifts for the state measurement probability. Because we do not know the direction of the shift, we assign systematic uncertainties to the branching fractions (See Table \ref{table:branching_fraction_systematics}).

We calculate the systematic shifts and uncertainties during state preparation due to the finite pumping time, decays from the $D_{3/2}$ and the $D_{5/2}$ states, and finite AOM extinction ratios by modeling the population evolution during state preparation for both pulse sequence (a) and (b). The measured pumping time constant for each state preparation step is $\leq \SI{1}{\micro\second}$, which is much shorter than the state preparation pulses that are $\geq \SI{50}{\micro\second}$. Therefore, the finite pumping time shifts the final branching fraction results by less than \SI{1e-9}{}, which is negligible compared to the statistical uncertainties, and therefore not included in Table \ref{table:branching_fraction_systematics}.

The finite state detection fidelity of the SD2 in sequence (b) and SD1 in sequence (a), due to Poisson statistics and the $D_{5/2}$ state decays, shifts the measured branching fractions. We calculate the shifts using the $D_{5/2}$ and $D_{3/2}$ state lifetimes and the $D_{5/2}$ branching fractions from Pal \emph{et al.} \cite{Pal2009}. In pulse sequence (b), step P2 pumps $D_{3/2}$ state population that has decayed from the $D_{5/2}$ state during SD2 to the $S_{1/2}$ state. Without P2, the pumping step P4 transfers residual population in the $D_{3/2}$  state to the $D_{5/2}$ state, and introduces a systematic uncertainty on the order of 1\% for $t$.

AOM light leakthrough could pump population to undesired states during state preparation. We measure the depopulation rates due to AOM leakthrough (see Appendix C), and calculate the systematic shifts and uncertainties due to AOM light leakthrough, which are included in Table \ref{table:branching_fraction_systematics}.

AOM leakthough of 802 nm during state measurements shifts the $D_{5/2}$ state population fractions, $p_{a}$ and $p_{b}$, calculated using the maximum likelihood method, as 802 nm leakthrough light also shifts the decay rate of the $D_{5/2}$ state. We add the 802 nm leakthrough depopulation rate to the $D_{5/2}$ state's natural decay rate, and use this total decay rate in the maximum likelihood model. The shifts and uncertainties are given in Table \ref{table:branching_fraction_systematics} under 802 nm AOM Leak SM.

Shifts due to off-resonant optical pumping are negligible for our measurements, as detunings between transitions are at least 50 THz. We determine the off-resonant pumping rate to be less than \SI{0.002}{\hertz} for our laser parameters, and the maximum uncertainty due to off-resonant pumping is more than two orders of magnitude smaller than the statistical uncertainty (see Appendix D), and thus not included in Table \ref{table:branching_fraction_systematics}.

All shifts are added linearly and uncertainties are added in quadrature for the final results in Table \ref{table:branching_fraction_systematics}. The systematic shifts and uncertainties are all smaller than the statistical uncertainties so they do not shift the statistical results significantly.
% ==============
\section{Theory}
% ==============
We evaluated the reduced matrix elements of the electric dipole $P_{3/2} -\, S_{1/2}$ and $P_{3/2} -\, D_{3/2,5/2}$
transitions in Ra$^+$ using the high-precision relativistic coupled-cluster single double triple (CCSDT) method~\cite{PorDer06}.
%developed in~\cite{BluJohLiu89,BluJohSap91,PorDer06}.
Ra$^+$ was considered as a univalent ion. We constructed the basis set in $V^{N-1}$ approximation (where $N$ is the number of electrons) in the framework of the Dirac-Fock approach, using 50 basis set B-spline orbitals of order 9 defined on a nonlinear grid with 500 points.
%Atomic units ($\hbar=|e|=m=1$) are used throughout.

These MEs were calculated previously in Ref.~\cite{Pal2009} in the framework of linearized coupled-cluster single double approximation.
In this paper we apply the more general CCSDT approach, additionally including valence triple excitations and non-linear (NL) terms into consideration.
The Breit interaction and quantum electrodynamic (QED) corrections were also taken into account.

The coupled cluster equations were solved in a basis set consisting of single-particle states. In the equations for singles and doubles the sums over excited states were carried out with 45 (of 50) basis orbitals with orbital quantum number $l \leq 6$.
The equations for triples were solved {\it iteratively}~\cite{PorDer06} but due to high computational demands we applied the following restrictions: (i) the core electrons excitations were allowed from the $[4s-6p]$ core shells,
(ii) the maximal orbital quantum number of all excited orbitals was equal to 3, and (iii) the largest principal quantum number $n$ of the virtual orbitals where excitations were allowed was 22.

The single-electron electric dipole moment operator, $\bf D$, is determined as ${\bf D} = -|e| {\bf r}$, where
$e$ is the electron charge and ${\bf r}$ is the radial position of the valence electron.
The reduced MEs $\langle S_{1/2} ||D|| P_{3/2} \rangle$ and $\langle D_j ||D|| P_{3/2} \rangle$ (in units of $|e| a_0$, where $a_0$ is the Bohr radius) are presented in~\tref{Tab:E1} and compared with other available data.

The results given on the line labeled ``LCCSD'' are obtained in the LCCSD approximation.
The lines 2-5 give different corrections. The corrections due to NL terms and valence triples are given on the lines 2 and 3.
On the lines labeled ``$\Delta$(Breit)'' and ``$\Delta$(QED)'' we present the Breit interaction and QED corrections, respectively. Both these corrections give a small contribution. For instance, the fractional contribution of the QED correction to the
$\langle S_{1/2} ||D|| P_{3/2} \rangle$ ME is only 0.12\%, which is in a good agreement with the value 0.14\% obtained in Ref.~\cite{RobDzuFla13}.
The final theoretical values are obtained as the sum of the LCCSD values and all corrections listed on the lines 2-5.

There are two main sources of uncertainty in the final theoretical values.
The first is due to an inaccuracy in the calculation of the correlation corrections,
and the second is due to an uncertainty of the QED corrections. The first uncertainty is estimated as the difference between the ``Final Th.''
and ``LCCSD'' values. The uncertainty of the QED corrections is estimated to be $\sim 25\%$. However, these corrections are small and their contribution to the uncertainty budget is negligible.

We note that our results are in very good agreement with the results obtained in the framework of the LCCSD approximation used
in Ref.~\cite{Pal2009}. As illustrated by \tref{Tab:E1}, the triple and NL corrections essentially cancel each other. Thus, such a
good agreement is not surprising. Our results are also in a good agreement with those obtained by~\citet{RobDzuFla13a} (where a
different approach based on correlation potential method~\cite{DzuFlaSil87} was used)
and with the results of Ref.~\cite{Sahoo2007}, where a similar, relativistic coupled cluster method, was applied.
%######################################################################################################
\begin{table}[tp]
\caption{Reduced MEs $\langle S_{1/2} ||D|| P_{3/2} \rangle$ and $\langle D_j ||D|| P_{3/2} \rangle$ (in $|e| a_0$). The values obtained in the LCCSD approximation and different corrections (see the main text for more details)
are presented. The final theoretical values (labeled as ``Final Th.'') are obtained as the sum of the LCCSD values
and all corrections listed on the lines 2-5. The uncertainties are given in parentheses.}
\label{Tab:E1}%
\begin{ruledtabular}
\begin{tabular}{lccc}
& \multicolumn{1}{l}{$\langle S_{1/2} ||D|| P_{3/2} \rangle$} & \multicolumn{1}{l}{$\langle D_{5/2} ||D|| P_{3/2} \rangle$}
& \multicolumn{1}{l}{$\langle D_{3/2} ||D|| P_{3/2} \rangle$} \\
\hline \\ [-0.6pc]
   LCCSD                 &  4.511     &  4.823     &  1.512     \\[0.1pc]
$\Delta$(NL)             &  0.056     &  0.080     &  0.028     \\[0.1pc]
$\Delta$(vT)             & -0.083     & -0.087     & -0.031     \\[0.1pc]
$\Delta$(Breit)          &  0.0002    & -0.011     & -0.002     \\[0.1pc]
$\Delta$(QED)            &  0.005     & -0.006     & -0.002     \\[0.3pc]
Final Th.                &  4.489(22) &  4.799(23) &  1.505(7)  \\[0.2pc]
Ref.~\cite{Pal2009}      &  4.511     &  4.823     &  1.512     \\[0.2pc]
Ref.~\cite{RobDzuFla13a} &  4.482     &  4.795     &  1.496     \\[0.2pc]
Ref.~\cite{SahWanJun09}  &  4.54(2)   &  4.83(8)&  1.54(2)
\end{tabular}
\end{ruledtabular}
\end{table}
%###################################################################################

%--------------------------------------------------------------
%\subsection{Branching ratios for decay of the $7p_{3/2}$ state}
%\label{7p3}
%--------------------------------------------------------------
Using the MEs given in \tref{Tab:E1} we are able to find the total decay rate of the $P_{3/2}$ state, $W_{\rm tot}$, and the branching ratios,
$r$, $s$, and $t$ (determined earlier), in different approximations. The total rate can be written as
the sum of the $P_{3/2} -\, S_{1/2}$ and $P_{3/2} -\, D_{5/2,3/2}$ transition rates, $W_{\rm tot} = W_r + W_s +W_t$, where
$W_r \equiv W(P_{3/2} \rightarrow S_{1/2})$, $W_s \equiv W(P_{3/2} \rightarrow D_{5/2})$, and $W_t \equiv W(P_{3/2} \rightarrow D_{3/2})$.
The probability of the $M1$ $P_{3/2}\, -\, P_{1/2}$ transition is negligibly small compared to the transition rates of other decay channels.

The results obtained in different approximations are given in the respective rows in \tref{Tab:Branch}. The CCSD results include NL terms but not triples, LCCSDT results include triples but not NL terms, and CCSDT values include both the NL terms and triples. ``Final Th.'' results for $r$, $s$, $t$, and $W_{\rm tot}$ are obtained as the sum of the ``CCSDT'' values and the Breit interaction and QED corrections.
%###################################################################################
\begin{table}[tp]
\caption{Branching fractions $r$, $s$, $t$, and $W_{\rm tot}$ (in $10^8$ $\SI{}{\per\second}$), obtained in different approximations. The final theoretical values (labeled as ``Final Th.'') are compared to the experimental results obtained in this paper (labeled as ``Expt.'') and previous theoretical results, Refs.~\cite{RobDzuFla13a,Pal2009,SahWanJun09}. The uncertainties are given in parentheses.}
\label{Tab:Branch}%
\begin{ruledtabular}
\begin{tabular}{lcccc}
&\multicolumn{1}{c}{$r$} & \multicolumn{1}{c}{$s$} &\multicolumn{1}{c}{$t$} & \multicolumn{1}{c}{$W_{\rm tot}$} \\
\hline \\ [-0.6pc]
LCCSD                    & 0.8768      &  0.1078     & 0.01541     &  2.116     \\[0.1pc]
CCSD                     & 0.8758      &  0.1086     & 0.01558     &  2.172     \\[0.1pc]
LCCSDT                   & 0.8768      &  0.1079     & 0.01534     &  2.049     \\[0.1pc]
CCSDT                    & 0.8757      &  0.1087     & 0.01553     &  2.093     \\[0.1pc]
Final Th.                & 0.8768(14)  & 0.1078(13)  & 0.01543(19) &  2.096(18) \\[0.2pc]
Expt.                 & 0.87678(20) & 0.10759(10) & 0.01563(21) &            \\[0.2pc]
Ref.~\cite{RobDzuFla13a} & 0.8767      &  0.1080     & 0.0153      &  2.089     \\[0.1pc]
Ref.~\cite{Pal2009}      & 0.8767      &  0.1078     & 0.0154      &  2.117     \\[0.1pc]
Ref.~\cite{SahWanJun09}  & 0.8773      &  0.1069     & 0.0158      &  2.142(42)
\end{tabular}
\end{ruledtabular}
\end{table}
%###################################################################################

The absolute uncertainty, $\Delta W_{\rm tot}$, of the total decay rate of the $P_{3/2}$ state is determined as
\begin{equation}
\Delta W_{\rm tot} = \sqrt{(\Delta W_s)^2 + (\Delta W_r)^2 + (\Delta W_t)^2} ,
\end{equation}
where the absolute uncertainties $\Delta W_s$, $\Delta W_r$, and $\Delta W_t$ are found using the uncertainties of the respective MEs
given in \tref{Tab:E1}. For calculation of the transition rates and branching fractions we use the experimental energies~\cite{DamJunWil16}
%\cite{RalKraRea11}
that are known with a high accuracy and do not contribute to the uncertainty budget.

Using the calculated MEs, we found the total decay rate of the $P_{3/2}$ state, the branching fractions, and their uncertainties. For instance,
the uncertainty of the branching fraction $r$ can be found using standard formulas from the equation,
\begin{eqnarray}
r = \frac{W_r}{W_{\rm tot}} = \frac{1}{1+ (W_s + W_t)/W_r},
\end{eqnarray}
and similar equations can be written for $s$ and $t$. Final theoretical values for $r$, $s$, and $t$ and their
uncertainties are presented in~\tref{Tab:Branch}. If the lifetime of the $P_{3/2}$ state is measured with a high precision, then using the experimental values for branching fractions we will be able to extract the values of the electric dipole MEs of the $P_{3/2} -\, S_{1/2}$ and $P_{3/2} -\, D_{3/2,5/2}$ transitions with a high accuracy.

There is very good agreement between the theoretical and experimental results. Using the MEs given in Refs.~\cite{RobDzuFla13a,SahWanJun09} and
the experimental energies we have calculated $r$, $s$, $t$, and $W_{\rm tot}$. These results, also presented in ~\tref{Tab:Branch} for comparison,
are in agreement with our values.

Using our calculations we are able to find the $s/t$ ratio and determine its uncertainty.
A standard formula to estimate the uncertainty of $s/t \equiv x$ is
$\Delta x = x\,\sqrt{(\Delta s/s)^2 +(\Delta t/t)^2}$ (where $\Delta x$, $\Delta s$, and $\Delta t$ are the absolute uncertainties of $x$, $s$, and $t$, correspondingly). Using $s= 0.1078(13)$ and $t= 0.01543(19)$ we obtain $\Delta x \approx 0.12$, noticeably overestimating the uncertainty because the formula for $\Delta x$ assumes that both quantities $s$ and $t$ change independently when we include different corrections. In reality the changes in these quantities, as seen from Table III, are essentially correlated. It is not surprising because we consider the transitions from the $P_{3/2}$ state to the fine structure states, $D_{3/2}$ and $D_{5/2}$.
Comparing the ``LCCSD'' and ``CCSD'' rows, we see that the NL corrections increase both $s$ and $t$ by 0.7 and 1.1\%, respectively.
The inclusion of triple corrections (cf. the ``LCCSDT'' and ``CCSD'' rows) decreases both $s$ and $t$ by 0.7 and 1.5\%, correspondingly. In such a way the errors in $s$ and $t$ values partially cancel each other in the ratio $s/t$ making it rather insensitive to different corrections.
For this reason we assume that it is correct to estimate the uncertainty of $s/t$ as the largest difference between the ``Final Th.'' and an intermediate (``CCSD'', ``LCCSDT'', or ``CCSDT'') value. We note that such an approach is not applicable to $W_{\rm tot}$, because in this case there is no mechanism for cancellation of errors in different terms and the standard method to determine its uncertainty should be used.

Using the final theoretical values of $s$ and $t$ and applying the method of estimating uncertainty discussed above we find the ratio $s/t = 6.990(43)$. This value is in a good agreement with the experimental result $s/t = 6.884(92)$ but is two times more accurate. Given the theoretical ratio of $s/t$ and the experimental high-accuracy value of $s=0.10759(10)$ we can extract the value  $t=0.01539(10)$.

\section{Summary}

\begin{figure}[H]
    \centering
    \includegraphics[]{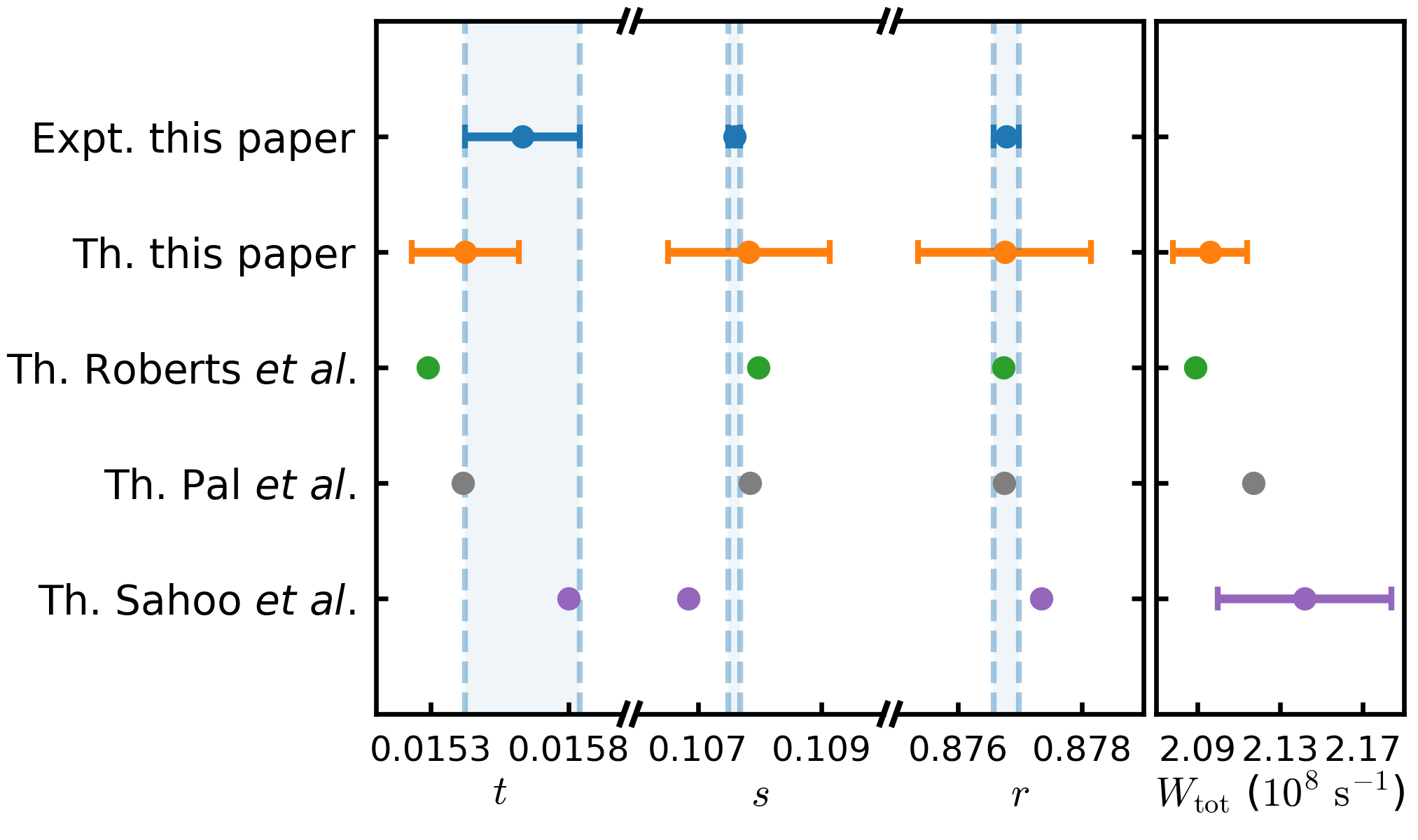}
    \caption{Comparison of the experimental results on branching fractions $r$, $s$, and $t$ and theoretical calculations from this paper, Th. Roberts \textit{et al.} \cite{RobDzuFla13a}, Th. Pal \textit{et al.} \cite{Pal2009}, and Th. Sahoo \textit{et al.} \cite{SahWanJun09}. Theoretically calculated values of the $P_{3/2}$ state total decay rates are shown.}
    \label{fig:theory_comp}
\end{figure}

Good agreement is found between measurements of the radium ion's $P_{3/2}$ branching fractions and our theoretical values, as well as with previous theoretical works (see Fig. \ref{fig:theory_comp}). The measurement precision of the $P_{3/2}$ branching fraction to the $S_{1/2}$ state supports a 0.1\% calculation of the $\langle S_{1/2} ||D|| P_{3/2} \rangle\langle D_{3/2} |H_{\mathrm{PNC}}| P_{3/2} \rangle$ PNC amplitude term, where $H_{\mathrm{PNC}}$ is the PNC Hamiltonian that mixes opposite parity states with the same total electronic angular momentum. This term accounts for $\sim6\%$ of the total PNC amplitude in a Ra$^+$ $S_{1/2}-D_{3/2}$ PNC experiment \cite{Pal2009}.  The measured $P_{3/2}$ branching fractions in this paper can be combined with Ra$^+$ light shift measurements to determine the $\langle D_{3/2} ||D|| P_{1/2} \rangle$ matrix element \cite{SahWanJun09}, which will improve the precision of the largest PNC contribution, the $\langle D_{3/2} ||D|| P_{1/2} \rangle\langle S_{1/2} |H_{\mathrm{PNC}}| P_{1/2} \rangle$ term \cite{Pal2009}.

\section*{Acknowledgement}

This research was performed in part under the sponsorship of the Office of Naval Research,
Grant No.~N00014-17-1-2252, NSF Grants No.~PHY-1912665 and No.~PHY-1620687, and the University of California Office of the President (Grant No. MRP-19-601445). S.P. acknowledges support by Russian Science Foundation under Grant No.~19-12-00157.

\appendix

\begin{table*}[ht]
\caption{Maximum off-resonant pumping rate of all relevant dipole transitions by lasers used in the experiment.}
\label{table:off_resonant_pump}
\begin{ruledtabular}
\begin{tabular}{lccccc}
%\rowcolor{gray!50}
    Light wavelength & $S_{1/2}-P_{3/2}$ & $S_{1/2}-P_{1/2}$  & $D_{3/2}-P_{3/2}$  & $D_{5/2}-P_{3/2}$ & $D_{3/2}-P_{1/2}$  \\
    \midrule
    468 nm & \SI{8e-7}{\hertz} & $\cdots{}$ & \SI{1e-4}{\hertz} & \SI{2e-5}{\hertz} & \SI{8e-6}{\hertz} \\
    708 nm & \SI{1e-7}{\hertz} & \SI{2e-7}{\hertz} & $\cdots{}$ & \SI{5e-4}{\hertz} & \SI{5e-5}{\hertz} \\
    802 nm & \SI{1e-7}{\hertz} & \SI{1e-7}{\hertz} & \SI{2e-3}{\hertz} & $\cdots{}$ & \SI{1e-4}{\hertz} \\
    1079 nm & \SI{6e-8}{\hertz} & \SI{7e-8}{\hertz} & \SI{3e-4}{\hertz} & \SI{1e-4}{\hertz} & $\cdots{}$ \\
\end{tabular}
\end{ruledtabular}
\end{table*}

\section{State measurement}

The 468 nm fluorescence during state measurement is uniformly distributed.   An ion that decays during state measurement from the $D_{5/2}$ state results in $(N_{\mathrm{b}}-N_{\mathrm{d}}) (1-t/t_{0}) + N_{\mathrm{d}}$ PMT counts on average, where $t_0$ is the state measurement time, $t$ is the time of decay, and $N_{\mathrm{b}}$ ($N_{\mathrm{d}}$) is the average number of PMT counts for bright (dark) events.  Because the state measurement time is much shorter than the $D_{5/2}$ state lifetime, $\tau_{5/2}$, we approximate the decay probability during state measurement as a constant, and therefore the total decay probability during state measurement for a dark event is $t_{0} / \tau_{5/2}$. The probability mass function of PMT dark event counts considering $D_{5/2}$ decays during state measurement is
\begin{equation}
\label{eq:dark_pmf}
\begin{aligned}
    D(N; N_{\mathrm{d}}, N_{\mathrm{b}}) = &(1 - t_{0} p_{\mathrm{decay}}) P(N; N_{\mathrm{d}})\\ &+ \int_{0}^{t_{0}} p_{\mathrm{decay}} P(N; (N_{\mathrm{b}}-N_{\mathrm{d}}) (1-t/t_{0}) \\&+ N_{\mathrm{d}}) dt
\end{aligned}
\end{equation}
where $P(N; N_{0})$ is the probability that a Poisson distribution with average value $N_0$ yields $N$, and $p_{\mathrm{decay}} = (\tau_{5/2})^{-1}$ is the $D_{5/2}$ state's decay rate.  The probability mass function for bright PMT counts is given by the Poisson distribution $ B(N; N_{\mathrm{b}}) = P(N; N_{\mathrm{b}})$.  We model the combined probability mass function for all PMT counts as

\begin{equation}
\label{eq:total_pmf}
    E(N; N_{\mathrm{d}}, N_{\mathrm{b}}, p_\mathrm{d}) = p_{\mathrm{d}} D(N; N_{\mathrm{d}}, N_{\mathrm{b}}) + p_{\mathrm{b}} B(N; N_{\mathrm{b}})
\end{equation}
where $p_{\mathrm{d}}$ is the state measurement probability for dark events, and for bright events $p_\mathrm{b}=1-p_\mathrm{d}$.

We use a maximum likelihood method to determine  $p_\mathrm{d}$  with Eq. \ref{eq:total_pmf}. To do this we maximize $\prod_i p_i$  by varying $N_{\mathrm{d}}$, $N_{\mathrm{b}}$, and $p_\mathrm{d}$, where  $p_i = E(N_i; N_{\mathrm{d}}, N_{\mathrm{b}}, p_\mathrm{d}) $  where $N_i$ are the PMT counts for the $i$th state measurement.  The parameters that maximize $\prod_i p_i$ are the maximum likelihood results.

With $p_\mathrm{d}$ and the total number of measurements, $M$, the dark state probability uncertainty is $\Delta p_{\mathrm{d}} = \sqrt{p_{\mathrm{d}}(1-p_{\mathrm{d}})/M}$  \cite{Gerritsma2008}.

\section{Collision rate}

We measure the collision rate using a pulse sequence with 1 ms state detection (which also cools the ion) after a 10 ms wait time. During the wait time the ion is left in the dark. If a dark state event is detected a collision event occurred. During 13500 s of measurement, we detected 23 dark events (consecutive dark events are counted as a single event, as either the ion was shelved in the $D_{5/2}$ state due to an inelastic collision or it was outside the imaging region due to a collision). This gives a collision rate of \SI{0.0017\pm0.0004}{\hertz}. We note that this collision rate measurement is also sensitive to electric-field noise heating, but the Doppler shift due to electric-field noise heating within the short duration of each experiment cycle is unlikely to affect state readouts.

\section{AOM leakthrough depopulation rates}

Leakthrough light due to finite AOM extinction drives radium ion transitions when the AOMs are off. We measure the depopulation rates of the lower states from the AOM leakthrough.

\textbf{468 nm: }
The 468 nm AOM leakthrough depopulation rate is measured using a pulse sequence that initializes the ion in the $S_{1/2}$ state, and then waits 50 ms before a 50-$\mathrm{\mu}$s-long 708 nm pulse. The 50-ms wait time is short compared to the calculated $D_{3/2}$ lifetime, 638(10) ms \cite{Pal2009}, so if population was shelved by AOM leakthrough light at 468 nm, the ion is most likely in the $D_{3/2}$ state at the end of the wait time. Light at 708 nm pumps 11\% of any population from $D_{3/2}$ to $D_{5/2}$ (from the $P_{3/2}$ branching fractions measured in this paper). After the 708 nm pulse, we state detect for 1 ms to measure the $D_{5/2}$ population. The measured depopulation rate from the $S_{1/2}$ state to the $D_{3/2}$ state by 468 nm AOM leakthrough light is \SI{0.0045\pm0.0010}{\hertz}.

\textbf{708 nm: }
We measure the 708 nm AOM leakthrough by initializing the population in the $D_{3/2}$ state and measuring the population in the $D_{5/2}$ state after a 2-ms wait time with state detection. The measured $D_{3/2}$ state depopulation rate  due to 708 nm AOM leakthrough light is \SI{0.005\pm0.002}{\hertz}. Because the wait time is short compared to the lifetimes of the $D_{5/2}$ and $D_{3/2}$ states \cite{Pal2009}, the shift of the depopulation rate due to either $D_{5/2}$ or $D_{3/2}$ decays during the 2-ms wait time is negligible compared to the statistical uncertainty of the depopulation rate.

\textbf{802 nm: }
We initialize the ion in the $D_{5/2}$ state, and measure its population after a 300-ms wait time first with the 802 nm light blocked by a mechanical shutter and second without the shutter.  Both data sets are fit to exponential decays. With the shutter the decay rate is  3.29(10) Hz and without the shutter the rate is 3.69(14) Hz.  This gives a $D_{5/2}$ depopulation rate of 0.40(17) Hz due to 802 nm AOM leakthrough.

\textbf{1079 nm: }
Measuring the 1079 nm AOM leakthrough is complicated by $D_{3/2}$ state decays. We initialize population in the $D_{3/2}$ state and then wait 100 ms before pumping a fraction of the population to the $D_{5/2}$ state  through the $P_{3/2}$ state.  By measuring the $D_{5/2}$ population after pumping we can infer the $D_{3/2}$ population at the end of the wait time. Depopulation due to 708 and 1079 nm AOM leakthrough, as well as spontaneous decays results in a $D_{3/2}$ state total decay rate of 1.90(20) Hz. This decay rate is greater than the spontaneous decay rate of 1.57(2) Hz from the theoretical natural lifetime \cite{Pal2009}. With the measured 708 nm depopulation rate and the $D_{3/2}$ state spontaneous decay rate, the 1079 nm AOM leakthrough depopulation rate is $\SI{0.33\pm0.20}{\hertz}$.

\section{Off-resonant pumping}

We analyzed the systematic effects due to off-resonant optical pumping.  Table \ref{table:off_resonant_pump} summarizes the off-resonant  pumping rates for all relevant dipole transitions assuming maximum light intensity at the ion (i.e., we assume the ion is centered in a Gaussian beam).  If we assume that all off-resonant pumping shifts the branching fraction values in the same direction we find that the shift is two orders of magnitude smaller than the statistical uncertainty. Therefore we do not include off-resonant pumping as a systematic uncertainty in
Table~\ref{table:branching_fraction_systematics}.

\end{document}